# Monolithic Integration of Piezo-Optomechanical Photonics and CMOS Electronics

Matthew Zimmermann[1,†], Aileen Zhai[2], Andrew J. Leenheer[3], Julia Boyle[1], Mayank Mishra[4], Daniel Dominguez[3], Matt Koppa[3], Wolf Jehle[3], Christopher Panuski[5], Mark Dong[1,5], Gerald Gilbert[6], Dirk Englund[5], and Matt Eichenfield[2,3,4,&]

*[1]The MITRE Corporation, 202 Burlington Road, Bedford, Massachusetts 01730, USA*

*[2]Electrical, Computer, and Energy Engineering, University of Colorado Boulder, Boulder, Colorado 80309, USA*

*[3]Sandia National Laboratories, P.O. Box 5800 Albuquerque, New Mexico 87185, USA*

*[4]Wyant College of Optical Sciences, University of Arizona, Tucson, Arizona 85719, USA*

*[5]Research Laboratory of Electronics, Massachusetts Institute of Technology, Cambridge, Massachusetts 02139, USA*

*[6]The MITRE Corporation, 200 Forrestal Road, Princeton, New Jersey 08540, USA*

[†] *mzimmermann@mitre.org*, [&] *matt.eichenfield@colorado.edu*

## Summary

Next-generation photonic architectures for AI, sensing, and quantum computing require thousands to millions of reprogrammable photonic devices on a chip[1]. The monolithic integration of Electronically-backed Photonic Integrated Circuits (EPICs) allows for very high density electrical interconnection and electronic drivers that can scale with photonics. Piezo-optomechanical photonic integrated circuits (POMPICs) offer low power consumption, high speed modulation, cryogenic compatibility and broadband optical transparency from ultraviolet to infrared wavelengths[2,3], but have not been demonstrated with monolithically integrated CMOS electronics. Here, we show a fully monolithic, all-CMOS fabricated platform for POMPICs co-fabricated with commercial control electronics. 200 millimeter photonic wafers are constructed directly on completed CMOS driver wafers by back-end-of-line processing, connecting integrated piezoelectric actuators under broadband silicon nitride waveguides to a high-density digital backplane comprising >2 million electrical connections per die with 6.4×6.4 micron electrode pitch. We introduce segmented POMPIC components as Photonic Digital-to-Analog converters (PDACs) that convert low-voltage digital electronic signals to multi-bit analog optical phase and amplitude modulation, and we demonstrate parallel control of optical phase shifters, Mach-Zehnder interferometers, optical routing trees, and tunable ring resonators using a standard HDMI interface to program CMOS electronics. We test multiple reticles and perform electronic and photonic characterization across the entire wafer to establish uniformity and yield, demonstrating wafer-scale integration of POMPICs on an electronic backplane and enabling dense, scalable electronic control of piezo-optomechanical circuits. This architecture provides a path to tightly integrated, high-channel-count control of quantum and classical piezo-optomechanical photonic integrated circuits, supporting applications in linear optical quantum computing, hybrid quantum systems with solid-state emitters, photonic neural networks, quantum networking, and display technologies.

## Introduction

Emerging photonic integrated circuit architectures show high potential when scaling to thousands of reprogrammable devices or more[1], but the need for electrical control that scale with photonic integrated devices is a growing concern for shrinking and scaling out photonic systems for quantum computing[4–6], neuromorphic photonic computing[7,8], large-scale optical phased arrays[9,10], and co-packaged optical transceivers or switches for chip-to-chip and system-to-system communication[11,12]. Size, weight, power, cost, latency, yield, and bandwidth are all critical concerns for the next generation of these systems, which require thousands of low power elements, broadband optical wavelengths, and tightly integrated electronic driving circuits. As the number of reconfigurable elements scale to the thousands and beyond, the co-integration of electronic drivers with photonic devices becomes crucial. Monolithic integration, which has been used to scale micro-electromechanical systems (MEMS)[13–15], radio frequency (RF)[16], and photonic[17] capabilities to higher bandwidths, large device counts, and lower latencies, offers an appealing roadmap[18] for scaling electronics alongside photonics for high density, high speed, and low latency interfaces between electrical drivers and integrated photonic devices. Foundry support exists for the monolithic integration of silicon photonics[19,20] but these are not compatible with the near-infrared (NIR), visible (VIS), or ultraviolet (UV) wavelengths that are needed for many quantum computing architectures, visible light displays, and biological imaging. Interposers using Through-Silicon Vias have enabled tightly integrated electronic-photonic packages of broadband Silicon Nitride ($SiN_X$) photonics[21], but these require extensive 3D assembly and processing. Furthermore, quantum computing schemes often require cryogenic operation[22,23], where tuning and control schemes of thermal modulation and carrier injection are infeasible when scaling to thousands or millions of devices[24]. Co-integrating electronic drivers with photonic devices in a monolithic process for broadband NIR, Visible, and UV wavelengths in foundry-compatible processes can support millions of electrical connections and photonic devices per die in inherently scalable processes, enabling new and powerful photonic architectures on a chip[25–27].

Piezo-optomechanical photonic integrated circuits (POMPICs)[28] have the potential to solve many of the gaps in the aforementioned photonic technology spaces. They offer high-speed, visible wavelength, cryogenically compatible (demonstrated down to <0.5K[29]) PICs, made in a CMOS-compatible process and demonstrating low- or zero- hold power and low drift for common photonic devices such as Mach Zehnder Interferometers (MZI's) and tunable ring resonators[2,30,31]. This platform also shows efficient resonant phase and frequency modulation[3], ultraviolet wavelength capabilities[32], and an architecture which allows for the heterogeneous integration of color centers in diamond[33,34] and other materials with $SiN_X$ waveguides. One of the reasons the platform is so versatile and powerful is that it can be fabricated in volume CMOS foundries, leveraging state-of-the-art tooling and modern materials. This offers the possibility of direct back-end-of-line (BEOL) fabrication of these piezo-optomechanical PICs on integrated driver circuitry. Challenges include lithographic alignment of the photonic layer with the underlying backplane, high temperatures needed for layer deposition, high electric fields in growing integrated piezoelectric layers, and a $XeF_2$ release process that can etch away upper layers of the PIC.

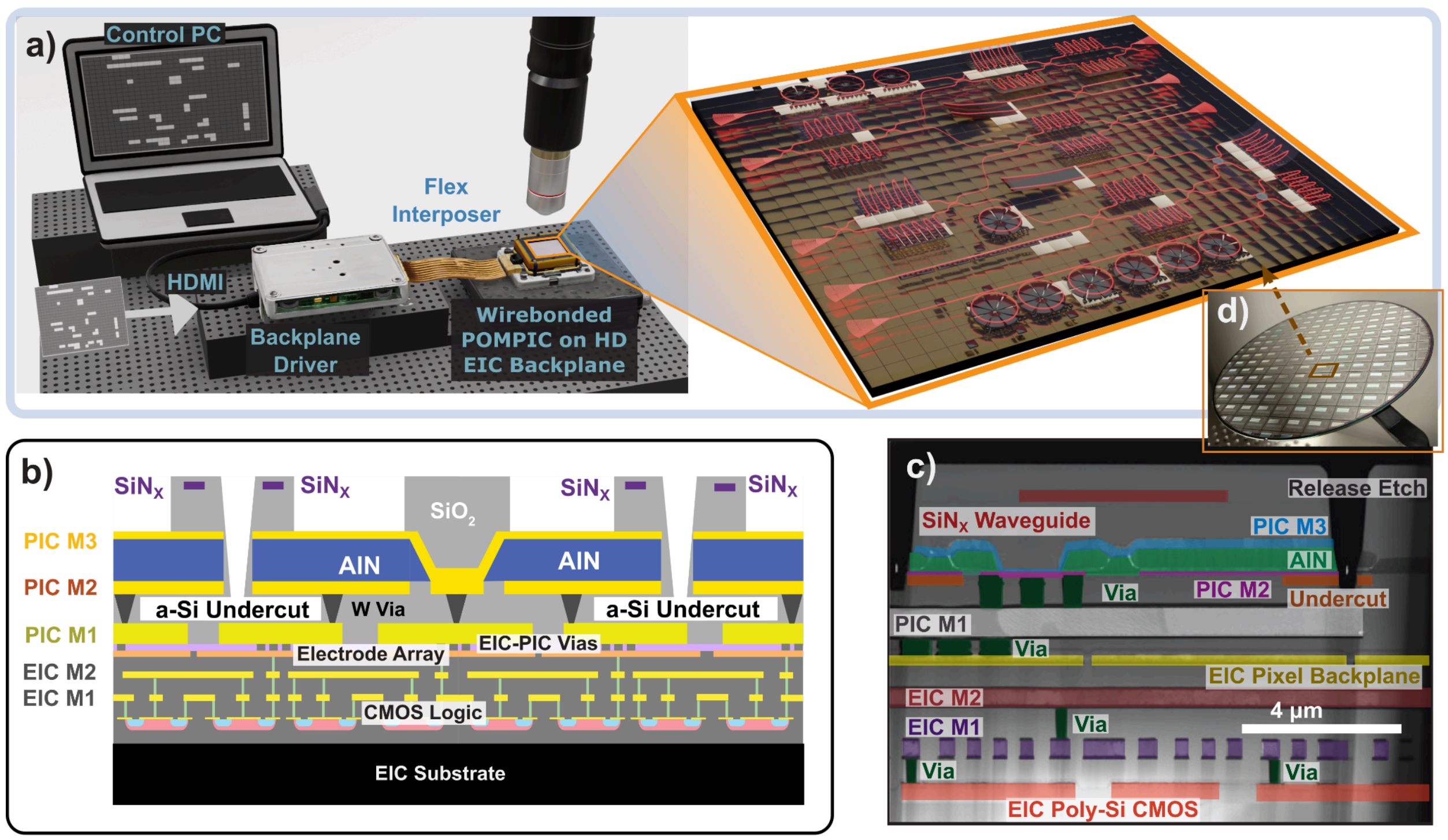

**Fig. 1 | CMOS-Backed Piezoelectrically Actuated Photonics. a,** Conceptual diagram of experimental setup. A control PC sends an image of 2 million actuatable segments as an HDMI image, which gets translated to high-speed data lines that drive segments on the CMOS backplane and actuates PIC devices. Zoom: Closeup of a conceptual CMOS-backed Piezo-Optomechanical PIC (POMPIC), with phase shifters, ring resonators, strain tuning elements, and integrated beamscanners, all based on prior work in the POMPIC platform. **b,** Layer stackup of the Piezoelectrically-actuated optomechanical PIC on CMOS Backplane. The HD backplane can independently drive both the top and bottom metal layers of the piezoelectric aluminum nitride. **c,** Focused Ion Beam (FIB) cross section of the PIC on EIC Backplane, showing the EIC layers, the vias which bridge the EIC to the PIC, and the Piezo PIC structure. False color has been added to show layers. **d,** Photograph of fabricated 200mm photonic integrated circuit wafer on an EIC backplane.

Heterogeneous integration (HI) is typically considered to have a yield advantage over monolithic integration of electronics and photonics, since the systems to be integrated can be separately verified before the HI processing. However, here we have shown that separately fabricated and fully qualified CMOS electronics wafers can be used as starting material for a monolithic, wafer-scale POMPIC fabrication process, completely eliminating these yield disadvantages. Here, we demonstrate an all-CMOS-fabricated POMPIC platform monolithically integrated with commercially manufactured CMOS driver electronics, with all fabrication performed in a volume CMOS foundry. The platform and processes are inherently scalable and are especially suitable for linear optical quantum computing[35,36], optical quantum control systems which require precise manipulation of optical fields[31], and hybrid approaches based on heterogeneous integration of solid state materials hosting solid state spin-photon interfaces[34,37]. We fabricate our photonic integrated circuit directly on top of a commercially available electronic wafer typically used for high-definition (HD) microdisplays[38], which has over 2 million available electrical connections per die via the HD backplane array. While this backplane has been used for die-bonding PICs[39–41], we demonstrate wafer-scale monolithic, back-end-of-line integration of a POMPIC with a multitude of photonic devices on this electronic backplane wafer. We design custom segmented versions of the POMPIC components that provide fractional phase shifts or resonance shifts when driven by a low-voltage digital backplane, creating visible-wavelength, piezoelectrically driven Photonic Digital-to-Analog Converters (PDACs)[42] which directly convert digital pulse-width-modulated drive signals from our CMOS backplane to optical phase- and amplitude- modulation with multiple bits of resolution. We demonstrate control of digital voltage signals to drive complex photonic circuit components on a digitally driven backplane with a 1920×1080 array of electrodes that connect to the top and bottom metal layers of our integrated aluminum nitride (AlN) piezoelectric actuators. This enables digitally driven bipolar high-resolution control of the phase of several different types of phase modulators, the amplitude of MZI's constructed from those phase modulators, and resonant detuning of ring resonators. These core components (phase shifters, MZI's, and ring resonators) are the building blocks for programmable photonic circuits[1], and the successful demonstration of scalable electronic integration is an important milestone for the POMPIC platform. We perform wafer level electrical and photonic testing to verify the entire wafer, and active testing demonstrating backplane-controlled integrated photonics across multiple reticles. This demonstration paves the way for scalable control of photonic integrated circuits for application to quantum computing, quantum sensing, photonic AI acceleration, and quantum networking applications.

## Results

The standard process for developing our POMPICs is described in our prior work[2,30]. While this unique layer stack is typically fabricated on a silicon substrate, for this work we replace the substrate with a 200 mm electronic CMOS wafer and begin our PIC fabrication with the first metal layer. For an electronic backplane, we chose the JD2552 2K voltage-drive backplane from Jasper Display Corp, a commercially available CMOS full digital backplane targeted for use in single and multi-channel optical systems for phase and amplitude modulation applications. One wafer contains 129 die instances, and each die contains a HD array of over 2 million independently-driven electrodes with 6.4 µm pitch. A vendor-supplied driver board converts a 60 Hz HDMI image to a high speed parallel data interface (64 bits at 300 MHz DDR), and on-chip logic updates the entire backplane. Within each HDMI frame, the backplane drives electrodes with low voltage (0-5V) pulse width modulated (PWM) signals that flicker at sub-millisecond rates. We design segmented versions of POMPIC devices as PDACs to take advantage of this dense digital distributed backplane, connecting dozens to hundreds of digital signals per modulator to independently drive the bottom (PIC M2) and top (PIC M3) metal around AlN actuators so each actuator can produce a phase shift of $-\delta\theta$, 0, or $+\delta\theta$ in our phase shifters and resonant detuning $-\delta f$, 0, or $+\delta f$ in our ring resonators due to the strain-optic effect, moving boundary effects, and path length changes in the $SiN_X$ waveguides. Driving multiple segments produces a total phase shift $\Delta\theta$ between arms of an MZI, or a total resonance shift $\Delta f$ when driving a ring resonator. MZI devices were lengthened from prior work so the total range of $\Delta\theta$ would be approximately $1\pi$ when going from a full negative shift to a full positive shift at 5V. Custom software is used to generate HD images to the vendor-supplied driver board to selectively drive electrodes in the backplane to produce target device actuation and demonstrate phase and resonance shifts.

Our fabrication process overcomes several key challenges that previously hindered monolithic CMOS integration of complex photonic platforms. Full detail of the POMPIC fabrication process is described in Methods and SI-1. For EIC-PIC interconnection, we use a plasma etch process to make small holes in the top $SiN_X$ passivation layer of the EIC wafer to allow access to the metal electrodes in the HD array. Subsequent CMOS-compatible fabrication processes 'fill in' holes with metal and create Vertical Interconnect Access (VIAs) between the digital backplane and the PIC. We add vias to EIC I/O signal pads at the edge of the chip to allow a high speed digital interface to program the entire electrode array through the EIC's on-chip logic. The fabrication process involves high temperatures around 400° C, high-power plasma processes for growing oriented aluminum nitride, and corrosive processes such as $XeF_2$ undercut requiring proper encapsulation of all transistors and waveguides. Damaging underlying CMOS or having a small angular misalignment of the first metal layer would short neighboring electrodes or fail to connect the backplane to the photonics, so lithography tools for PIC layers were precisely aligned to the EIC backplane. We verify that the underlying EIC was not damaged and properly aligned by measuring expected electrical behavior across the entire wafer through photonic and electrical probing (SI-2), and we confirm backplane-driven piezo-optomechanical driving of three different wirebonded dies without any excessive current draw or any evident loss in CMOS functionality after adding the PIC layers, and alignment pads on the top and bottom of the electrode array allow us to directly probe the signal on backplane electrodes and correlate the backplane driving signal with our photonic response (SI-3).

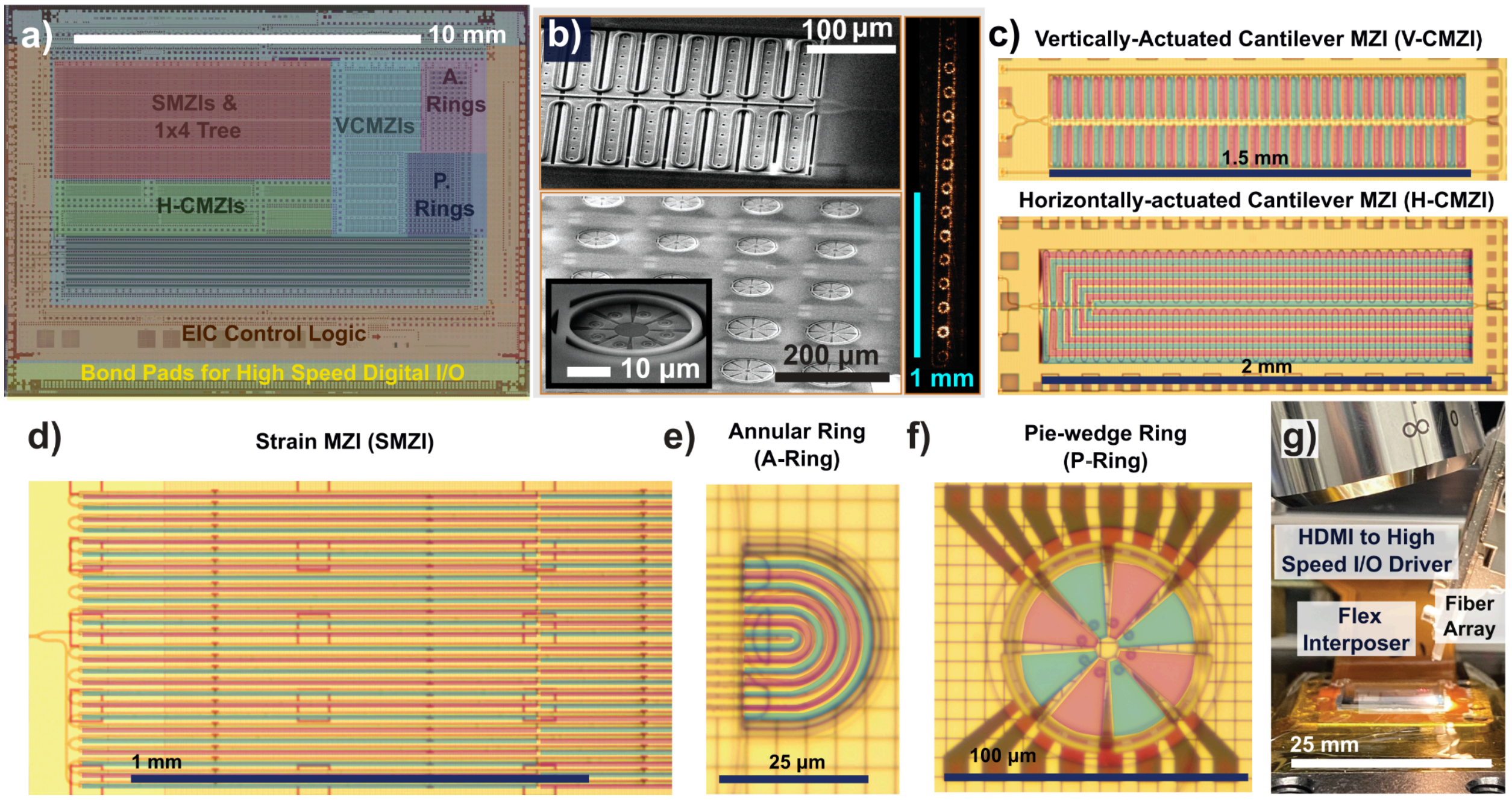

**Fig. 2 | Fabricated CMOS-Backed PIC. a,** Photograph of the fabricated PIC on CMOS Backplane, with PIC devices labelled. **b,** Scanning Electron Microscope (SEM) images of V-CMZI (top) and Pie-Wedge Rings (lower) after $XeF_2$ Release. Right: Top-down microscope photo of chained P-Ring device showing scattered light. **c-f,**. Microscope photos of **(c, upper)** V-CMZI, **(c, lower)** H-CMZI, **(d)** SMZI, **(e)** Annular Rings, and **(f)** Pie Wedge Rings. False color (alternating pink and green) has been added to show discrete segments of the piezoelectric actuators tied to electronic backplane; each segment connects two electrodes in the EIC backplane to the bottom and top metal layers of our integrated piezoelectric actuators. **g,** Photograph of experimental setup in lab

The completed reticle is shown in Fig 2. We designed various MZI's using different POMPIC phase shifting mechanisms; these include Strain Optic-based MZI's (SMZI's), Vertically-Actuated Cantilever MZI's (V-CMZI's), and Horizontally-actuated Cantilever MZI's (H-CMZI's). We also include a large array of optical ring resonators capable of optomechanical strain detuning by actuating the piezoelectric layer underneath using two actuator types: Pie Wedge Actuated Rings (P. Rings) and Annular Actuated Rings (A. Rings). Focused Ion Beam (FIB) cross section analysis on the SMZI, shown in Fig. 1c, was performed to show the full layer structure of the active device, including vias that connect the EIC to the backplane of digitally driven electrodes. Here, we observe that the underlying electronics layer was preserved; the $XeF_2$ release did not penetrate into the EIC, the vias are well aligned with the HD backplane, and the EIC vias, routing metal, and transistor layer are visible and do not show signs of cracking or damage from the high temperatures, pressures, charged electrostatic fields, or $XeF_2$ release involved in POMPIC fabrication. Fig 2b shows SEM images of a VCMZI and P. Rings after the devices have been released, as well as a chained P. Ring device. To allow us to measure the electrical signal directly, we route vias at the top and bottom of the electrode array to bond pads at the top of the PIC so we can directly measure the electrical signal from the backplane while setting target electrodes. Temperature monitoring of a wirebonded die with an FLIR camera showed the temperature of the chip changed by only 7.6° C when driving all electrodes with no active cooling (Supplemental SI-4).

Fig 1a and Fig 2g show the experimental setup for measuring modulation from the backplane. Because the backplane outputs a digital PWM signal, this produces a flickering response. To show how the piezoelectric components perform and demonstrate successful backplane-driven actuation, we measure the high-speed PWM modulation on high speed photodiodes and measure the settled state. More information on the experimental setup is in Supplemental SI-3.

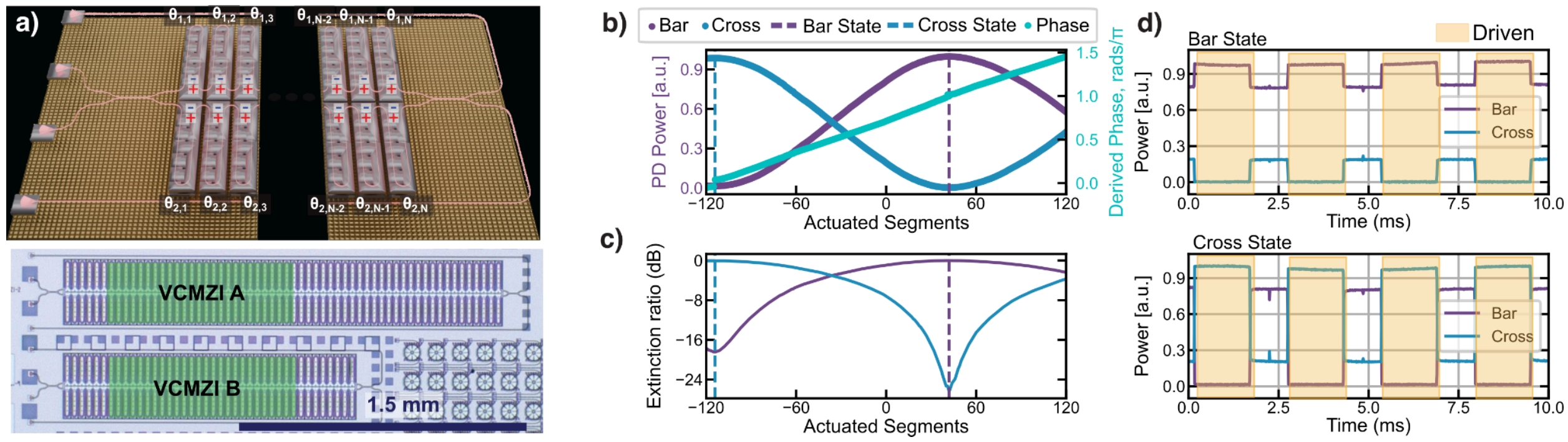

**Fig. 3 | Backplane-Controlled Vertically-Actuated Cantilever MZI (V-CMZI). a,** Top: Computer Generated Model of device. Each phase shifter is made up of a series of actuators with one waveguide loop per actuator. When digitally driven, each segment can give a small positive phase shift or a small negative phase shift. Bottom: Microscope photo of two V-CMZI devices, with $N_A$=60 and $N_B$=40 segments per arm. **b,** Measured output power and derived phase shift from a V-CMZI A device at 778 nm. Power is measured as the settled value when N segments are driven. **c)** Measured extinction ratio in dB. **d,e)** High-speed actuation of device in its bar (d) and cross(e) states. Shaded region shows when device is driven to its target state; non-shaded region shows device in its undriven state.

The first device we characterize is the V-CMZI, shown in Fig. 3a. The device is segmented so each waveguide loop is a separate actuator, and there are N segments in both the top and bottom arm, providing an actuation range from -2N to +2N. We design devices with N=40 and N=60, and with 5V actuation, we observe consistent phase shift of 0.02 radians/segment, requiring a range of 160 segments for a full π phase shift. This gives us both high precision and a high tuning range, with digitally driven photonics providing a measured extinction ratio of >20 dB and a voltage-length product of approximately 25 Vπ · cm, which is consistent with prior work in piezo-actuated photonic cantilevers. We measure a rise time of approximately 15 µs – this is consistent with the capabilities of the backplane driver, and a dedicated high-speed driver would achieve faster actuation times.

The second device we characterize is the segmented Strain-optic MZI (SMZI), shown in Fig. 4. This device, a segmented version of the device characterized in prior work[2], applies strain under segments of widened waveguides. These devices have N=104 segments in both the top and bottom arms, allowing for a discrete tuning range of ±208. At 5V actuation, we observe a consistent phase shift per segment of 0.009 radians, and require a range of 340 segments for a full π phase shift. We measure an extinction ratio of >30 dB on a single device,Because strain-based phase shifting is not as effective as cantilever phase shifters, the voltage-length product is higher (170 $V\pi \cdot$ cm) but is consistent with prior work[2]. We note the rise time of the SMZI devices is limited due to the electronic backplane, but also note that piezoelectronic components with similarly sized actuators have been shown to actuate at much faster speeds of 100 MHz-1 GHz[2,43]. We also demonstrate this device as a 1x4 SMZI Switching Tree, and demonstrate the device pulsing between four target outputs (Fig. 4c), showing that this architecture is compatible with larger scale trees, circuits, and meshes.

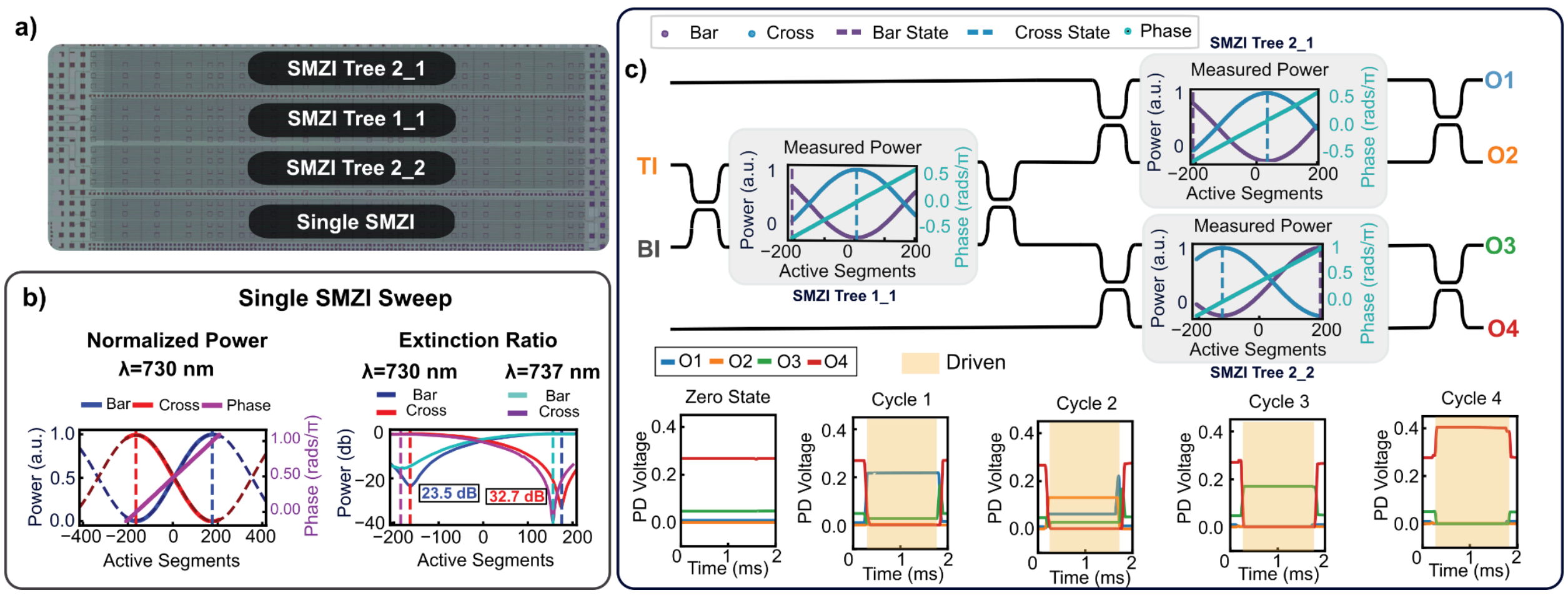

**Fig. 4 | Backplane-Controlled Strain-optic MZI. a,** Microscope Image of SMZIs. The SMZI tree is in a switchback configuration, so all optical inputs and outputs are on the left side. **b,** Left: Measured cross and bar state power levels (normalized) and derived phase of Single SMZI $\Delta\theta$ at $\lambda$=730 nm. Dashed curve shows fitted sinusoidal response. Right: Extinction ratios shown for $\lambda$=730 and 737 nm. **c,** Block diagram of 1x4 SMZI tree and experimentally measured results of the cross and bar state power levels. Lower: Switching demonstration of the 1x4 tree. HDMI output is periodically updated to alternate between the 4 different output states. Shaded region shows when device is driven to its target state; non-shaded region shows device in its undriven state.

The third device we characterize is the horizontally-segmented Cantilever MZI (H-CMZI). This device, shown in Fig 5, uses N=8 L-shaped segments per arm to perform phase shifts. This gives a full tuning range of ±16, but as the outer segments are larger than inner segments, we expect and observe a less linear response, with 'center' segments providing no distinguishable phase shift but other segments providing larger shifts. We show results from 2 devices of varying lengths, and observe phase shifts ranging from 0 to 0.2 radians. This progressive phase shift could enable higher-precision tuning with fewer segments by tailoring future devices to have binary-like shifts, with outer segments giving coarse tuning and inner segments giving fine tuning to precisely hit target phase. We note that the best performing device, H-CMZI A, measured phase shifts that ranged from 0.07 to 0.15 radians (SI-6), with mean phase shifts of 0.11 radians/segment and a voltage length product of 24 $V\pi \cdot$ cm.

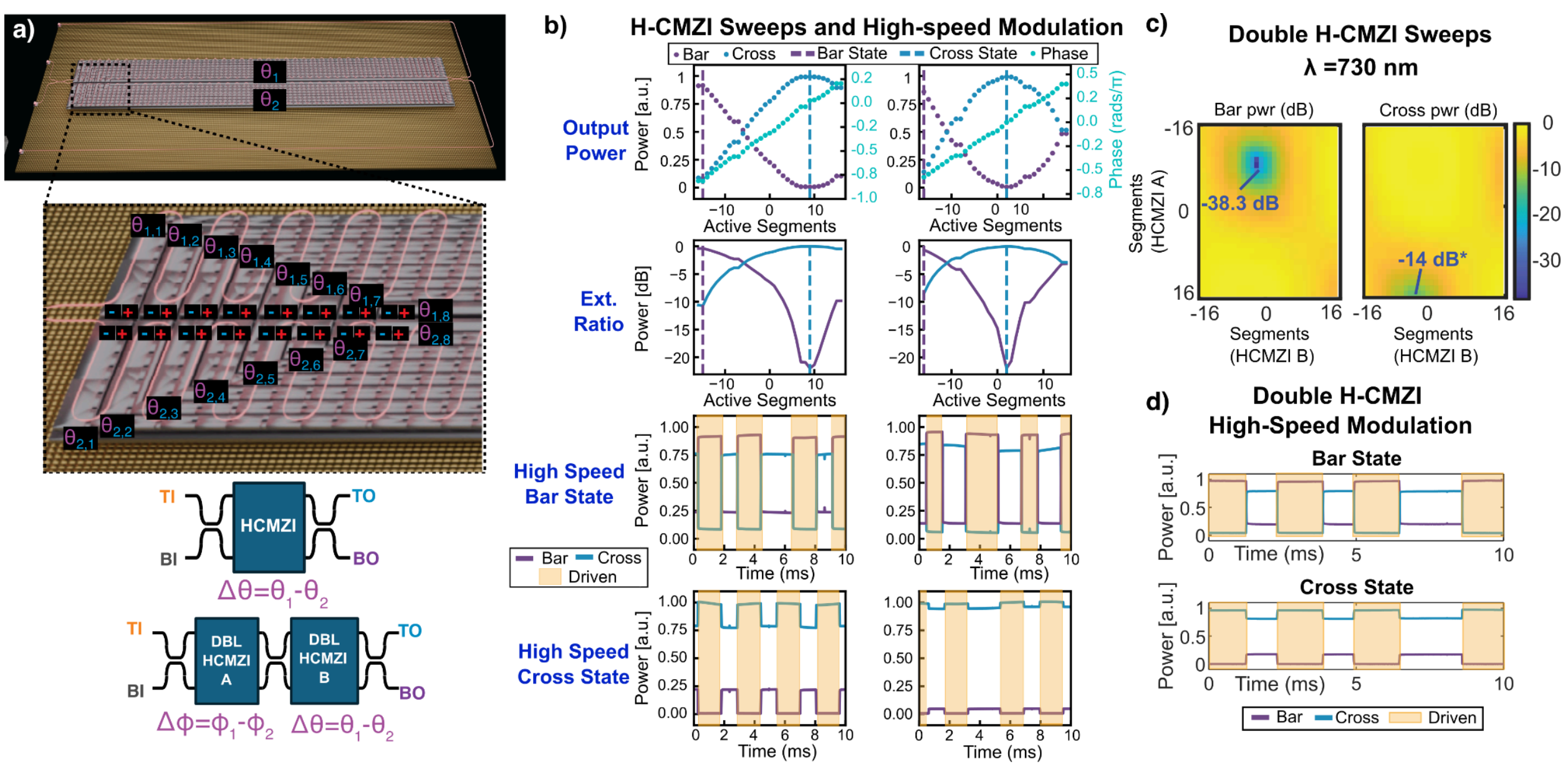

**Fig. 5 | Backplane-Controlled Horizontal CMZI. a,** Computer Generated Model of H-CMZI. Each segment forms an 'L'-shaped actuator that stretches across the full width of the cantilever. An MZI with three beamsplitters and two phase shifters allows for high extinction ratios that can correct for beamsplitter errors. **b**, Backplane Sweep and High Speed Data for Horizontal-MZI. (Left: H-CMZI A, L=2250, O=140; Right: H-CMZI B, L=2800, O=140). Shaded region shows when device is driven to its target state; non-shaded region shows device in its undriven state. **c,** 2D response of double-MZI, showing 35 dB extinction ratio of bar state. True cross state was slightly out of the tuning range of the device. **d,** High-speed data of device being toggled between the undriven states and the bar/cross states.

To correct for imperfect beamsplitting ratios, we demonstrate using the H-CMZI in a triple-beamsplitter configuration[44]; this uses three beamsplitters and two phase shifters. We do a 2D raster scan of actuated segments and measure bar power extinction of 38.3 dB. While we were unable to configure the device in its full cross state, modeling shows that with a higher tuning range we will approach a similar extinction ratio of >30 dB (SI-7).

**Table 1 | Backplane-Controlled MZI Characterization**

| Device | Discrete Tuning Range | Path Length (cm) | λ (nm) | Total Phase Shift Δθ (rads) | Shift per segment δθ (rads) | Segments for π Phase Shift | Vπ · cm | Bar State ER (dB) | Cross State ER (dB) | Rise Time (us) | Capacitance per segment (pF) |
|---|---|---|---|---|---|---|---|---|---|---|---|
| V-CMZI A | -120 to +120 | 3.8 | 780 | 4.71 | 0.020 | 160 | 25.2 | 18.5 | 26.4 | 15 | 1 |
| V-CMZI B | -80 to +80 | 2.5 | 780 | 3.14 | 0.020 | 160 | 25.2 | 11.7 | 31.0 | 15 | 1 |
| H-CMZI A | -16 to +16 | 2.4 | 730 | 3.14 | 0.07 - 0.15 (avg: 0.10) | 32 | 24.0 | 22.0 | 10.7* | 15 | 6 |
| H-CMZI B | -16 to +16 | 3.0 | 730 | 3.14 | 0.09 - 0.19 (avg: 0.12) | 32 | 30.0 | 21.9 | 8.5* | 15 | 7 |
| SMZI | -208 to +208 | 20.8 | 730 | 3.85 | 0.009 | 340 | 170 | 23.5 | 32.7 | 40 | 2 |

*Actual bar/cross state was out of device's tuning range

The results of the Mach Zehnder Interferometers are tabulated in Table 1. V-CMZI devices show consistent performance, good extinction, and fast actuation, with extremely low capacitance actuators. H-CMZI's have similar voltage-length products, and SMZI's give the finest tuning accuracy. To characterize these devices as PDACs that convert digital electrical signals to analog phase and power modulation, we use the measured phase shift per segment $\delta\theta$ to calculate the phase resolution of our phase shifters in bits as $B_{\theta} = log_{2}(\frac{\pi}{\delta\theta})$. We find phase resolution levels of 8 bits for SMZI phase shifters, 7 bits for V-CMZI phase shifters, and 5 bits for H-CMZI phase shifters in their current configuration.

Using the device for amplitude modulation as an MZI, the output power is proportional to the phase difference θ between MZI arms as $P_{out}(\theta) \propto sin(\frac{\theta}{2})^{2}$. As this is a nonlinear response, we measure output power resolution based on the sharpest response of the output power at $\theta = \pi/2$ to calculate the output power bit resolution of these devices as PDACs as $B_{PWR} = -\ log_{2}(0.5\ -\ sin(\frac{1}{2}\ (\frac{\pi-\delta\theta}{2}))^{2})$. From this, we show output power resolution levels of 7 bits for the SMZI, 6 bits for the V-CMZI, and 4 bits for the H-CMZI. Higher resolution levels can be achieved through analog voltage levels or tailored actuator design, and grouping segments can deliver binary phase- or amplitude- shifts. We note that all devices showed connectivity to the backplane, and we tested devices on three separate reticles from the fabricated wafer, all of which showed a functional electronic backplane and responsive piezo-optomechanical phase shifters. While these devices were not designed for low optical loss, we estimate the implementation loss in the V-CMZI's to be approximately 3 to 4 dB/cm after comparing the measured output powers of the two devices. This loss is comparable to prior work in this platform, and recent fabrication improvements to the same BEOL POMPIC processes show losses as low as 0.3 dB/cm for high-confinement waveguide devices and 0.03 dB/cm for low-confinement waveguide devices[45].

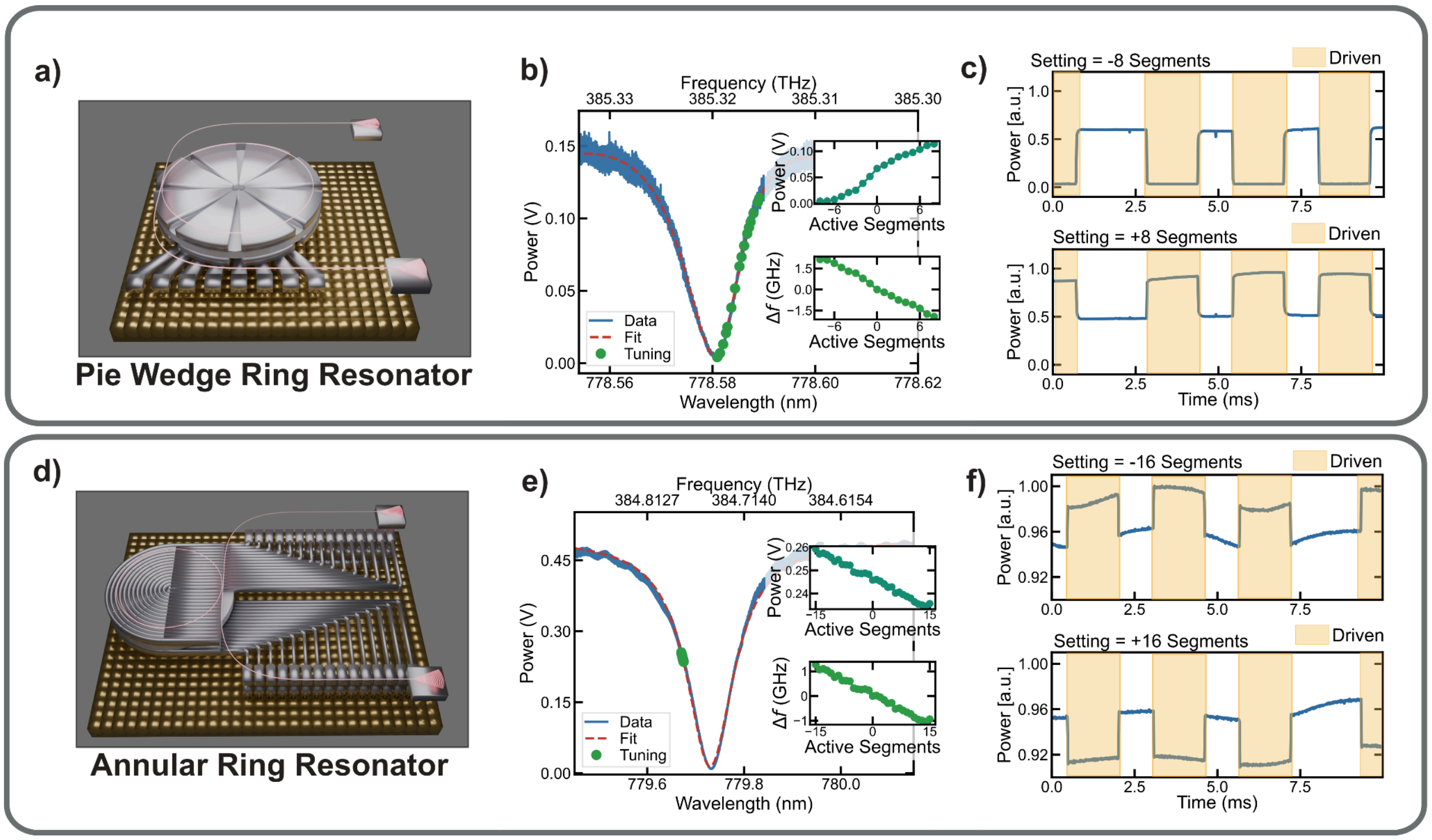

**Fig. 6 | Optical Ring Resonators on an integrated backplane. a,** Computer Generated Model of Pie Wedge Ring Resonator (P-Ring). **b,** Wavelength Sweep and backplane modulation of pie wedge ring resonator. Green points show the settled optical output photodiode level when the backplane was modulated. Upper Inset: Settled optical output measurement from photodiode. Lower Inset: Derived ring detuning in picometers. **c,** High speed modulation data. Top: Setting all 8 segments to a negative resonant shift to decrease amplitude. Bottom: Setting all 8 segments to a positive resonant shift to decrease amplitude. **d,** Computer Generated Model of Annular Ring Resonator (A-Ring). **e,** Wavelength Sweep and backplane modulation of annular ring resonator. Green points show the settled optical output photodiode level when the backplane was modulated. Upper Inset: Settled optical output measurement from photodiode. Lower Inset: Derived ring detuning in picometers. **f,** High speed modulation data. Top: Setting all 16 segments to a negative resonant shift to increase amplitude. Bottom: Setting all 16 segments to a positive resonant shift to decrease amplitude.

In addition to MZI's, we test two configurations of optical ring resonators. The first configuration, the Pie Wedge ring (P-Ring), shown in Fig. 6a-c, uses actuators of equal size underneath a ring. The entire ring is undercut, and actuating a segment will put strain on one slice of the ring. Actuating the AlN layer with an applied voltage puts strain on the $SiN_X$ waveguide in the ring, causing a shift in the ring's resonance. Each ring has 8 segments, and each segment connects to two electrodes, allowing for a positive or negative resonant shift. We first characterize the ring by sweeping the wavelength with a tunable laser, then set the laser to be tuned to the 50:50 point before modulating the segments of the backplane. In Fig. 6b, we observe a change in amplitude which we can map to resonant wavelength detuning. We find detuning to be fairly linear, and show an average tuning of 0.5 pm/segment (250 MHz/segment).

The second configuration, the Annular ring (A-Ring), shown in Fig. 6d-f, uses a series of U-shaped actuators to apply strain on the edge of the ring opposite to the waveguide. The ring is undercut to create an actuatable cantilever under the ring. We observe a change in amplitude while setting the pixel backplane, though the small actuator size and the small Q factor of the rings limits the tuning amplitude. We observe roughly linear wavelength detuning of 0.1-0.23 pm/segment (40-115 MHz/segment), and we note that outer segments produce a more noticeable shift compared to inner segments.

**Table 2 | Backplane-Controlled Optical Ring Resonators**

| Device | Discrete Tuning Range | Radius (um) | Extinction (dB) | FSR (THz) | Q | Total Shift $\Delta f$ (GHz) | $\Delta f$/V (MHz/V) | Shift per segment $\delta f$ (MHz) | Rise Time (us) | Capacitance per segment (pF) |
|---|---|---|---|---|---|---|---|---|---|---|
| P-Ring 1 | -8 to +8 | 40 | 15.1 | 0.63 | $6.71 \times 10^4$ | 4.08 | 408 | 255 | 15 | 0.11 |
| P-Ring 2 | -8 to +8 | 30 | 12.9 | 0.84 | $5.01 \times 10^4$ | 2.56 | 256 | 160 | 15 | 0.11 |
| A-Ring 1 | -16 to +16 | 30 | 7.4 | 0.84 | $6.22 \times 10^3$ | 2.08 | 208 | 65 | 15 | 0.05 |
| A-Ring 2 | -16 to +16 | 30 | 9.9 | 0.84 | $6.51 \times 10^3$ | 2.55 | 255 | 80 | 15 | 0.05 |
| A-Ring 3 | -16 to +16 | 30 | 18.9 | 0.84 | $6.94 \times 10^3$ | 2.95 | 295 | 92 | 15 | 0.05 |
| A-Ring 4 | -16 to +16 | 30 | 18.9 | 0.84 | $6.94 \times 10^3$ | 2.46 | 246 | 77 | 15 | 0.05 |
| A-Ring 5 | -12 to +12 | 25 | 13.1 | 1.0 | $7.93 \times 10^3$ | 2.05 | 205 | 85 | 15 | 0.04 |
| A-Ring 6 | -12 to +12 | 25 | 20.3 | 1.0 | $7.91 \times 10^3$ | 2.66 | 266 | 111 | 15 | 0.04 |
| A-Ring 7 | -12 to +12 | 25 | 24.4 | 1.0 | $7.49 \times 10^3$ | 1.92 | 192 | 80 | 15 | 0.04 |
| A-Ring 8 | -9 to +9 | 15 | 26.5 | 1.7 | $5.56 \times 10^3$ | 1.65 | 165 | 92 | 15 | 0.03 |

**$\lambda \cong 780$ nm for all devices.**

The results of ring resonators are tabulated in Table 2. Full wafer-scale analysis of the ring resonators is in SI-5, and detailed analysis of active die-level testing is in SI-8. We show that the backplane is capable of tuning the rings, and we divide the total measured tuning range by 10 volts to get the measured phase shift per volt when actuating the entire ring ($\Delta f/V$), and we find this is consistent with prior work. The rise time and fall times are in line with the speed of the backplane, and the extremely low capacitance per segment implies these devices could be driven at GHz-rate speeds and extremely low power with a custom electronic backplane.

Overall, these results successfully demonstrate high density, reliable electrical interconnect from an integrated high-density electronic backplane to a novel integrated photonic platform. We show high speed PWM modulation and precise tuning, with measured phase shifter performance $V\pi \cdot cm$ and ring detuning $\Delta f/V$ is consistent with prior work in the POMPIC platform[2,30,46]; therefore we do not see a degradation in photonic performance from fabricating POMPIC devices on an active electronic backplane instead of a blank silicon substrate. Our wafer scale probing (SI-2, SI-5) shows high yield and reliability across the entire wafer. Taken as a whole, these results provide a successful demonstration of processes for Electronically-backed POMPICs with commercial electronic backplanes, and this successful demonstration provides a roadmap for scaling novel POMPIC integrated circuits and applications.

## Discussion

In this work, we have demonstrated a process for wafer-scale monolithic CMOS integration of POMPICs on a commercial CMOS electronic backplane, which uses 120 electrical connections to drive 2 million independent control signals at a 6.4×6.4 µm pitch at a 60 Hz update rate via 300 MHz CMOS logic, column and row drivers, and a pulse width modulated backplane. We demonstrate integrated control of key devices for photonic information processing and applications: broadband phase shifters and Mach Zehnder modulators as well as ring-resonator-based tunable filters and resonant amplitude/phase modulators. This is a direct solution to the scaling problem of electronic control of large-scale NIR, VIS, and UV PICs and demonstrates the feasibility of wafer-scale co-integration of electronic backplanes with the low-power, low-loss, cryogenically compatible (<0.5K demonstrated)[29], optically broadband POMPIC platform. Our process couples electronic drivers to photonic devices directly above the driving circuit, eliminating the need for impedance control and reducing heat dissipation, and is compatible with a variety of CMOS process nodes for alternative commercially available CMOS IC's available for driving LED arrays[47] and MEMS devices[48] or custom Application-Specific Integrated Circuit development. Low load capacitance and minimal wiring between drivers and devices reduce parasitic effects and allow for low power, high speed piezoelectric modulation, and our digitally-driven devices demonstrate PDACs with multi-bit resolution, allowing ICs to provide analog optical control of phase and amplitude via binary logic[49], analog drivers[50], or both. Finally, our process leaves the PIC wafer's waveguide layer exposed and available for optical access, hybrid and heterogeneous integration, and post-fabrication processing such as $XeF_2$ release of piezoelectric actuators, integrated lasers[51], integration of color centers in diamond[34], quantum dot arrays[52], microdisk resonators[53], photonic crystal cavities[43], superconducting nanowire single photon detectors[54], low-loss cryogenically compatible optical packaging[29], and lithium niobate integration for 100 GHz-rate modulation[55].

Our method shows consistent performance and no degradation of the underlying EIC or POMPIC devices, and we foresee many potential paths for improvement. Stress engineering and device design can improve effective strain shift[43], resonant drive of piezo-optomechanical devices greatly enhances actuation[56], and actuation can be further enhanced by using scandium-aluminum nitride[57]. While this work shows moderate optical loss of approximately 2 to 4 dB/cm, fabrication improvements to the same BEOL POMPIC processes show losses as low as 0.3 dB/cm for high-confinement waveguide devices and 0.03 dB/cm for low-confinement waveguide devices[45]. Other $SiN_X$ platforms have demonstrated even lower losses and ultra-high-Q microresonators[58]. To reduce the size of our devices, compact Dual-Ring Mach Zehnder Modulators (DRMZM's)[31] provide compact, scalable, high fidelity light control, and combining capacitive electrostatic and piezoelectric tuning, recently demonstrated in this POMPIC platform[59], has shown a way to broadly tune cantilever phase shifters via capacitive pull-in, achieving capacitive $V_\pi$ voltages below 5 volts through DC Biasing. EIC processes such as Bipolar-CMOS-DMOS[60] allow for high voltage drivers and smaller photonic devices, while lower voltage electronics and larger photonic devices can be used for power-sensitive applications. With dedicated electronic drivers, this can be developed into a Process Development Kit (PDK) with Parameterized Cells (P-Cells) for scalable co-integrated EIC-PIC's that are cryogenically compatible, low power, broadband, high speed and scalable, and the digitally driven control scheme we demonstrate with accurate, multi-bit phase and amplitude modulation of visible wavelength optical signals provides an approach for digital backplanes to provide high-resolution analog control. The capabilities that this integration enables are far ranging and impactful for many applications that benefit from large arrays of reconfigurable broadband photonic devices.

Fault-Tolerant Quantum Computing (FTQC) architectures, which often require tightly scheduled low-latency processing of thousands of optical and electrical signals near cryogenic Quantum Processing Units, are often constrained at the electrical level[61,62]. Many proposed architectures are limited by power consumption, thermal budget[63], and electrical and optical I/O[62]. Combining a UV-VIS-NIR PIC with cryo-CMOS electronics[61,64,65] and large-scale fabrication and integration of solid state emitters[66] and detectors allows the unit cells of a potential quantum computer and their tailored electronic drivers to scale out at the foundry level, providing a scalable roadmap for FTQC and >1000 qubit architectures based on light and light-matter interactions[5,22,62].

Outside the cryogenic environment, tightly integrated beamscanning systems using laser beam scanning[67] or optical phased arrays[68] that can cover the UV-VIS-NIR spectrum are important for quantum control systems[31,69], displays[70], NIR LiDAR[71], advanced microscopy[72,73], 3D printing[74], and communication[75,76]. Arrays achieve increased precision, beam count, and range as the number of antenna elements increases[77], but electronics, wiring, and power consumption limits the maximum number of elements in state-of-the-art systems. The POMPIC platform's integrated AlN layer also allows for scalable acoustic control for beamsteering[78] or quantum control[79].

The same electronic control constraints we have addressed here also constrain the potential of photonic information processing architectures[50], optical circuit switching[80], and neuromorphic computing[81], which can vastly reduce the power consumption in AI, ML, and datacenter applications but require large meshes of stable reconfigurable photonic devices and integrated transceivers and photodetection. Our approach, which combines low-power piezoelectric tuning, broadband silicon nitride (and can easily be adapted to alumina for blue/UV photonics), a tightly integrated electronic backplane, and an open waveguide layer amenable to heterogeneous integration of other materials – all built in foundry-scalable processes – allows the number of reconfigurable elements to scale to the thousands or millions and reduces the size, cost, weight, and power of next-generation photonic systems.

## Methods

### Device fabrication and I/O

The fabrication process follows the approach laid out in Stanfield et. al[30], with the key difference of using a wafer with prefabricated commercial electronic integrated circuits instead of a bare silicon substrate. The commercial digital backplane wafer consisted of 180-nm node CMOS with Al routing up to an array of 6.4x6.4 micron Al electrodes coated with thin SiO2 and SiNx passivation. For PIC fabrication, the wafer was first coated with additional SiO2 and polished, then vias were etched through the SiO2/SiNx/SiO2 stack to contact the Al electrodes and filled with tungsten using a damascene polish process. Alignment of these vias to the HD electrode array on the backplane and subsequent photolithography layers was accomplished with careful offsets measured on each wafer at each layer for this split-fab process using different stepper lithography tools. A cross-section schematic is provided in Fig. 1b; after the via connection to the EIC electrodes, the first layer deposited during PIC fabrication is PIC M1 (Al). After an additional $SiO_2$ deposition, an amorphous silicon (a-Si) layer is deposited below the cantilever. The next layers are $SiO_2$, Al, AlN, Al, $SiO_2$, $Si_3N_4$, and lastly $SiO_2$. A defined etch around the cantilever exposes the underlying a-Si. After wafer fabrication is complete, the PICs are placed in a xenon difluoride ($XeF_2$) gas etching chamber which etches away the a-Si, releasing actuatable devices such as cantilevers except for a clamp at one end. The standard $Si_3N_4$ waveguides used on these devices were 300 to 400 nm wide and 300 nm thick. There is a $SiO_2$ cladding buffer between 1.0 and 2.6 μm on the left and right sides of the waveguide(s), along with a target top $SiO_2$ clad thickness of ~700 nm and a target bottom $SiO_2$ clad thickness of ~750 nm.

Pads for electrical contact to the control signals in the backplane use routing metal and tungsten vias to independently route signals to the EIC backplane. Signals from the EIC backplane route up through metal and tungsten vias to the top and bottom electrodes of piezoelectric actuators in the photonic devices. Electrical signals are sent to these devices from the electrode array on the backplane, which is programmed from a vendor supplied board which converts HDMI data from custom-generated HD images to PWM control signals that drive the electrode array. Laser light is routed to and from devices using a grating-coupled fiber array.

## Acknowledgements

Major funding for this work is provided by MITRE for the Quantum Moonshot Program. D.E. acknowledges partial support from Brookhaven National Laboratory, which is supported by the U.S. Department of Energy, Office of Basic Energy Sciences, under Contract No. DE-SC0012704 and the NSF RAISE TAQS program.

M.E. performed this work, in part, with funding from the Center for Integrated

Nanotechnologies, an Office of Science User Facility operated for the U.S. Department of Energy Office of Science.

Sandia National Laboratories is a multimission laboratory managed and operated by National Technology & Engineering Solutions of Sandia, LLC, a wholly owned subsidiary of Honeywell International Inc., for the U.S. Department of Energy's National Nuclear Security Administration under contract DE-NA0003525. This paper describes objective technical results and analysis. Any subjective views or opinions that might be expressed in the paper do not necessarily represent the views of the U.S. Department of Energy or the United States Government.

We thank H. Laroque for insights into developing multi-mode interferometers on a pixel backplane. A.J.L and W.J. thank C. Ozuna for fabrication assistance to align the PIC fabrication patterns to the underlying commercial wafer. We thank C. Brabec, S. Trajtenberg, and M. El-Kabbash for their insights into using the Jasper Display Technology. M.Z. thanks M. McBride, S. Vergados, Y. Kidane, and T. McManus for their work in developing electronic test systems and wirebonding the EIC-PIC. We thank G. Clark, A. Witte, T. Olsson, and M. Storey for valuable feedback on the manuscript.

## Author contributions

M.Z. supervised the project planning and progress, developed backplane compatible versions of PIC components, designed the test plan, took experimental data, processed data, and wrote the majority of the manuscript. A.Z. took experimental data, performed data processing, developed the conceptual models and figures, performed ring resonator analysis, and assisted with manuscript preparation. J.B. developed the control software and performed experimental measurement, data processing, and assisted with writing the manuscript. C.P. and D.E. proposed the concept of digital backplane wafers for photonic control and acquired the wafers used in this work. A.J.L. oversaw device fabrication, managed the foundry process, and provided technical guidance for PIC design. W.J. performed lithography alignment measurements to enable fabrication. D.D. designed PIC devices and helped oversee device fabrication and autoprober testing. M.K. performed photonic autoprober testing and data processing. M.M. processed the photonic autoprober data and experimental data and assisted with writing the manuscript. M.D. assisted in PIC component design and reticle preparation. M.E. conceived the PIC platform and developed the fabrication architecture with A.J.L., provided technical guidance and insights, coordinated institutions, and assisted with writing the manuscript. M.E., G.G., and D.E. supervised all aspects of the project.

## Competing interests

D.E. is a Scientific Advisor to and co-founder of Axiomatic_AI. D.E. is also a scientific advisor to and holds shares in QuEra Computing. The other authors declare no competing interests.

## Data availability

The data that support the plots within this paper are available under restricted access due to MITRE's information security policies. Access can be obtained from the corresponding authors upon request.